\documentclass[amsmath,showkeys,superscriptaddress,onecolumn,11pt]{revtex4}
\usepackage{float}
\usepackage{titlesec}
\usepackage{amsmath}
\usepackage{amsmath}
\usepackage{graphicx}
\usepackage[colorlinks=true,linkcolor=blue,citecolor=blue,urlcolor=blue,breaklinks]{hyperref}\providecommand{\keywords}[1]{\textbf{\textit{Index terms---}} #1}
 
\usepackage{graphicx}
\usepackage[colorlinks=true,linkcolor=blue,citecolor=blue,urlcolor=blue,breaklinks]{hyperref}

\usepackage{amssymb}
\begin{document}
\title{Probing tripartite entanglement and coherence dynamics in pure and mixed independent classical environments}
\author{Atta Ur Rahman}
\email{Attazaib5711@gmail.com}
\address{Department of Physics, University of  Malakand, Khyber Pakhtunkhwa, Pakistan}
\author{Muhammad Javed}
\author{Arif Ullah}
\address{Department of Physics, University of  Malakand, Khyber Pakhtunkhwa, Pakistan}
\begin{abstract}
Quantum information processing exploits non-local functionality that has led to significant breakthroughs in the successful deployment of quantum mechanical protocols. In this regard, we address the dynamics of entanglement and coherence for three non-interacting qubits initially prepared as maximally entangled GHZ-like state coupled with independent classical environments. Two different Gaussian noises in pure and mixed noisy situations, namely, pure power-law noise, pure fractional Gaussian noise, power-law noise maximized and fractional Gaussian noise maximized cases are assumed to characterize the environments. With the help of time-dependent entanglement witnesses, purity, and decoherence measures, within the full range of parameters, we show that the current mixed noise cases are more detrimental than pure ones where entanglement and coherence are found short-lived. The power-law noise phase, in particular, appears to be more flexible and exploitable for long-term preservation effects. In contrast, we find that in both pure and mixed noise cases, where entanglement and coherence degrade at a relatively high rate, there is no ultimate solution for avoiding the detrimental dephasing effects of fractional Gaussian noise. The three-qubit state becomes disentangled and decoherent within independent classical environments driven by both pure and mixed Gaussian noises, either in long or short interaction time. In addition, due to the lack of the entanglement revivals phenomenon, there is no information exchange between the system and the environment. The three-qubit GHZ-like states have thus been realized to be an excellent resource for long enough quantum correlations, coherence, and quantum information preservation in classical independent channels driven by pure power-law noise with extremely low parameter values.
\end{abstract}
\maketitle
\section{Introduction}
Information processing and computing based on quantum mechanics are the focus of Quantum information processing \cite{01, 02, 03}. Quantum computers are not restricted to two states, unlike current digital computers, which encode data in binary digits. Quantum bits, or qubits, are used to encode information and can exist in various types of superposition states. To act as computer memory and processors, qubits can be created using atoms, ions, photons, or electrons, as well as control devices \cite{04}. These computers offer a significant advantage over classical computers, as they can simultaneously hold multiple states. This empowers currently available algorithms, such as integer factorization or quantum many-body system simulation, to solve certain problems much faster than a classical computer \cite{05, 06, 07}. As a result, quantum computers have emerged as forefront advanced devices that are much superior in operation and most certain in practical applications with high speed and accuracy to their classical equivalents \cite{1,2,3}. Various quantum phenomena, that control the tasks to be performed are the key concerns and working concepts of quantum computers \cite{4,5,6,7,8}. The phenomena along with other principles of quantum computing are just as essential to consider as quantum computing devices are \cite{9,10,11,12}. Among many other non-local phenomena, entanglement and coherence are a few of the most important for establishing efficient quantum mechanical operations \cite{13,14,15}. Without a question, entanglement and coherence are at the core of almost all quantum mechanical processing protocols. The successful deployment of these protocols needs the useful transfer and preservation of such non-local correlations and coherence during realistic quantum operations. Quantum correlations dynamics of different quantum systems took precedence over the research problems in quantum information sciences \cite{16,17}.\\
Entanglement is a non-local phenomenon where two or more particles are intrinsically submerged into a single inseparable system. This was recognized as a key factor in distinguishing between local and non-local correlations \cite{18,19}. Hence, the entanglement derived remarkable attention and is investigated for novel non-local operations such as quantum communications \cite{20}, quantum cryptography \cite{21}, quantum dense coding \cite{22,23}, quantum teleportation \cite{24}, quantum secure direct communication and quantum key distribution \cite{25,26,27,28}. Any quantum computation that cannot be done efficiently on a classical computer requires maintainable entanglement \cite{29,30}. Besides, coherence is also at the core of quantum computation where qubits are superimposed in the states of "0" and "1," leading to acceleration over many classical algorithms. Quantum coherence is a prerequisite for entanglement and other types of quantum correlations, and it can be wielded to preserve entanglement in a quantum system.\\
Engineering, building, and programming quantum computers is extremely complex. They are hampered by errors such as noise, malfunctions, quantum decoherence and quantum disentanglement \cite{030, 031, 032}. This becomes critical to their functioning and breaks down before any nontrivial program can complete. This is due to the limitation because a quantum system cannot be kept isolated from the effects of the linked environments \cite{31,32,33}. The interaction of quantum systems with their surroundings, which results in decoherence and entanglement degradation, is becoming one of the major obstruction for the successful employment of quantum information processing protocols \cite{34,35,36}. This weakening or disappearance of the entanglement is known as disentanglement and is caused by decoherence. This environmental defect, where the initial state entanglement of the system is constrained, is due to environmental noises. It is therefore very important to optimize and characterize the evolution of the quantum systems in presence of such fatal interfering noises in particular, for practical utilization of quantum information processing. Environmental noise can be understood into two different categories, which are classical and quantum interaction pictures. The classical description is more important than the quantum equivalent since, it allows for a greater number of degrees of freedom to explore the time evolution of quantum systems \cite{37,38,39,40,41,42,43}.\\
Investigation of the time-evolution of entanglement, coherence, and their protection has been studied extensively and has remained the major focus in research aspects for various types of quantum systems. In addition, for both bipartite and tripartite quantum systems, different non-Gaussian noises such as random telegraph, static, and coloured noises are studied in detail \cite{37, 38, 53, rr, ii}. The findings show that these noises have dephasing effects that cause non-local correlations, coherence, and quantum information to be lost in a non-monotonic fashion, rendering the quantum process inefficient.. Important measures and methods are introduced to characterize the dynamical characteristics, coherence and entanglement dynamics of various types of quantum systems. In this regard, the multi-qubit systems have been showed with more cryptographic behaviour, and more superior in carrying information than the single-qubit and two-qubit systems \cite{44,45,46,47}.\\
This paper aims to investigate the dynamics of entanglement and coherence for a system of three non-interacting qubits initially prepared as maximally entangled Greenberger-Horne-Zeilinger ($\mathcal{X}_{GHZ}$) state in pure and mixed independent classical noisy environments. These environments are further assumed to be described by power-law (PL) and fractional Gaussian (FG) noise \cite{48,49,50}. The FG noise is caused by the discrete Brownian motion of the medium's particle and has been vastly researched in the case of traffic controls, electrical measurements, and meteorological data \cite{KAPLAN, Murad, FGn}. Multi-scale patterns, Tsallis permutation entropy, empirical mode decomposition, creating self-similar network traffic, and other hydrological issues have all been modelled using FG noise \cite{pentland, plastino, Rilling, leDEsma, Kout}. The low frequency PL noise spectrum expands as $\frac{1}{f^\alpha}$, where $f$ is the cyclic frequency and $\alpha$ is a real number. Thermal fluctuations and resistance, voltages across diodes and in vacuum tubes, and nearly in every solid-state device, frequency variations in harmonic fluctuations, and voltages in most superconducting devices are all sources of this noise \cite{Dutta, Postma, Kasdin, Kasdin-2, Lakhmanskiy}. We focus on the independent coupling of the three non-interacting qubits with three individual environments. We believe that independent system-environment coupling is more important than common and other types of environments because, in most cases, quantum subsystems are coupled with separate environments rather than a common one. This independent system-environment coupling is further extended to be described by pure PL, pure FG, power-law noise maximized (PLM) and fractional Gaussian noise maximized (FGM) configuration. The first two cases are related with classical environments with only pure noise. The latter cases are related to mixed noise situations where the effects of a Gaussian noise are maximized relative to other environmental noise. This is achieved by increasing the noise parameter values and assigning more number of qubits to a particular noise. Different measures, such as entanglement witness, purity, and decoherence will evaluate the dynamics of the tripartite entanglement and coherence in the presence of the current pure and mixed noisy situations.\\
The paper is organized as: In Sec.\ref{$QC$ measures}, the estimators used to measure tripartite entanglement, purity, and environmental decoherence are illustrated. Sec.\ref{The Model}, presents the physical model that accounts for the system-environment interaction, and the application of the pure and mixed noisy configurations. Sec.\ref{Results and Discussions}, deals with the results obtained for our physical model, as well as the discussions that followed. Sec.\ref{Conclusion}, represents the conclusive comments based on the investigation carried out.
\section{Tripartite entanglement and coherence measures}\label{$QC$ measures}
This section describes the quantifiers used to evaluate tripartite entanglement, decoherence.
\subsection{Entanglement witness}\label{dd}
Entanglement witnesses, which are Hermitian operators with at least one negative eigenvalue, facilitate experimental detection of entanglement \cite{EW1, EW2, EW3}. Entanglement witnesses exist as a result of functional analysis's Hahn-Banach theorem, which provides a necessary and sufficient condition for detecting entanglement \cite{EW4}. Mathematically, entanglement through this operation is computed as \cite{52}:
\begin{equation}
E(t)=-Tr[\mathbb{W}_{o} \rho_{abc}(t)]\label{ewo},
\end{equation}
where $\mathbb{W}_o=\frac{1}{2}\textit{I}-\rho_o$ with $\rho_o$ being the initial density matrix defined as $\rho_o=\vert \psi\rangle \langle\psi\vert$ and $\rho_{abc}(t)$ is the time evolved density matrix of the state $\psi$. For $E(t)=0$, the state will be separable. Negative results of the Eq.\eqref{ewo} for tripartite states suggest a high entanglement regime, but positive results may show both separability and a weak entanglement regime. Here, we will use the entanglement witness to identify the state with the true tripartite entanglement of the GHZ class that is not biseparable.
\subsection{Purity}
One of the most basic quantifiable measures of quantum coherence is purity. Within Gaussian noise-driven classical fields, it will estimate the degree of mixedness and loss of coherence of the pure coherent state in time. For a time evolved state ${\rho_{abc}(t)}$, purity is given by \cite{53}:
\begin{equation}
P(t)=Tr[\rho_{abc}(t)]^2. \label{purity}
\end{equation}
For $n$-dimensional quantum system, the purity criteria ranges as $ \frac{1}{n} \leq P(t) \leq 1$. The state is completely pure and coherent at $P(t)=1$, while becomes completely mixed and decoherent at the lower bound value $\frac{1}{n}$.
\subsection{Decoherence}
Decoherence occurs when quantum system's wave functions become entangled with their coupled environments. As a result, instead of being a single coherent quantum superposition, the system behaves like a classical statistical ensemble of its constituents. Decoherence will be a valid measure to compute coherence loss in the time evolved state of the system because the system-environment interaction is described classically here. The Von-Neumann entropy approach can be used to estimate the decoherence effects for the time evolved density matrix $\rho_{abc}(t)$ as \cite{53}:
\begin{equation}
D(t)=-Tr[\rho_{abc}(t)ln\rho_{abc}(t)].\label{decoherence}
\end{equation}
For entangled and coherent quantum states, $D(t)=0$, whereas any other value of this measure will indicate the corresponding amount of coherence loss.
\section{The Model}\label{The Model}
Our physical model comprising three identical non-interacting qubits with equivalent energy splitting $\varepsilon_n$ that are exposed to external independent classical random fields. We assume these fields to be characterized by Gaussian statistics, which further are presented in pure and mixed noisy configurations. 
\begin{figure}[ht]
\includegraphics[width=13cm,height=12cm,keepaspectratio]{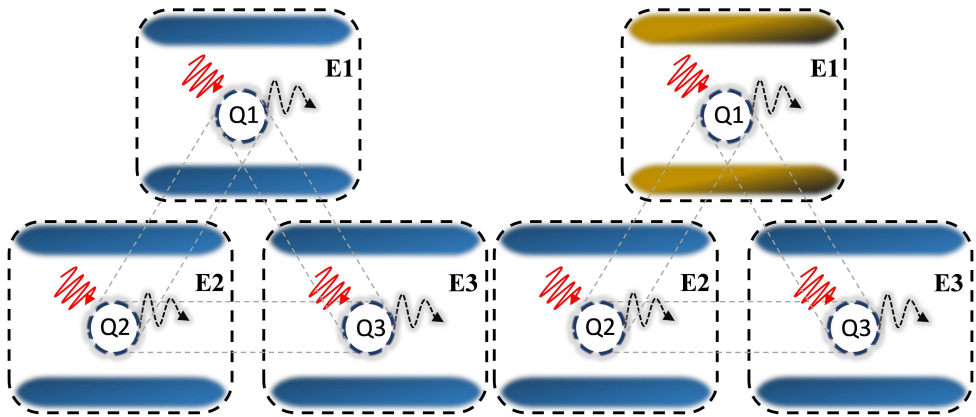}
\caption{The Schematic diagram of three non-interacting qubits $Q_1$, $Q_2$ and $Q_3$ coupled with independent classical environments $E_1$, $E_2$ and $E_3$ represented by square like boxes. The model with single-coloured regions represents the action of pure Gaussian noise (left) while the configuration with two coloured regions shows the action of mixed Gaussian noises (right) i.e. power-law noise maximized configuration and fractional Gaussian noise maximized configuration. The black wavy lines represent the dynamics of the system with reduced amplitudes means the dephasing effects of the noise in the corresponding subspaces of the qubits. The red wavy lines indicate the action of the noises with the connecting lines among the qubits reflect the non-local correlations between the subsystems and showing that the three qubits are prepared as a single composite state. The identical size and shape of the qubits show that they are assembled with equal energy splitting.}
\end{figure}
In the first case, we investigate the dephasing effects because of pure PL and FG noises. In a mixed noise situation, we consider power-law noise maximized (PLM) and fractional Gaussian noise maximized (FGM) configuration. In the PLM configuration, two qubits are coupled with PL noise and one with FG noise. Similarly, in the FGM case, two qubits are coupled with FG and one with PL noise.
The current physical set-up is ruled by the stochastic Hamiltonian, written as \cite{53}:
\begin{equation}
H_{123}(t)=H_1(t) \otimes I_2 \otimes I_3+I_1 \otimes H_2(t) \otimes I_3+I_1 \otimes I_2 \otimes H_3(t),\label{hmm}
\end{equation}
where $ H_n(t)$ is the single qubit Hamiltonian and is defined by $ H_n(t)=\varepsilon_n I_n+\lambda \Omega_n(t)\sigma_n^x$ with $n \in \{1,2,3\}$. Here, $\Omega_n(t)$ is the stochastic parameter randomly flipping between $\pm1$ while $\lambda$ is the coupling constant. $I_n$ and $\sigma_n^x$ are the identity and Pauli matrices acting on the sub-spaces of the qubits. The time evolution of the system can be obtained by \cite{rr}:
\begin{equation}
\rho_{123}(t)=U_{123}(t)\rho_{GHZ}(0) U_{123}(t)^{\dagger},\label{Ham}
\end{equation}
where, $U_{123}(t)$ is the time unitary operator and is defined as $U_{123}(t)=\exp[-\int^{t}_{t_o}H_{123}(s)ds]$ with $\hbar=1$. Here, $\rho_{GHZ}(0)$ is the initial density matrix of the system and is given as \cite{dd}:
\begin{equation}
\rho_o = \frac{I_{(8\times 8)}(1-r)}{8}+r\vert \mathcal{X}_{GHZ}\rangle\langle \mathcal{X}_{GHZ}\vert,
\end{equation}
where $r \in \{0,1 \}$ and $\mathcal{X}_{GHZ}$ is the three qubit maximally entangled Greenberger-Horne-Zeilinger state and can be written as $ \mathcal{X}_{GHZ}=\frac{1}{\sqrt{2}}(\vert 000 \rangle + \vert 111 \rangle)$.
\subsection{Fractional Gaussian and power-law noise}
The thorough application of the FG and PL noises will be discussed in this section. In order to include the stochastic process in the case of classical noises, the $\beta$-function must be specified, which introduces the noise phase to the system and reads as \cite{55,56}:
\begin{equation}
\beta_n(t)=\int_0^t \int_0^t dz dz^\prime k(z-z^\prime).\label{beta function}
\end{equation}
The auto-correlation functions of the noises, which relate the noise phase to the phase of the system in the local fields are written as:
\begin{equation}
J_{FG}(t-t^{\prime})=\frac{1}{2}(|t|^{2H}+|t^{\prime}|^{2H}-|t-t^{\prime}|^{2H}),\label{auto-correlation fn of FN}%
\end{equation}
\begin{equation}
J_{PL}(t-t^{\prime},\chi,\Gamma,\alpha)=\frac{\alpha-1(\chi \Gamma)}{2(\chi \vert t-t^{\prime}\vert+1)^2}.\label{auto-correlation fn of PL}
\end{equation}
Now, upon assuming the dimensionless noisy quantities $g=\frac{\chi}{\Gamma}$ and $\tau=\Gamma t$ and inserting the auto-correlation functions from Eq.\eqref{auto-correlation fn of FN} and \eqref{auto-correlation fn of PL} into Eq.\eqref{beta function}, the corresponding $\beta$-functions are obtained as:
\begin{equation}
\beta_{FG}=\frac{\tau^{2H+2}}{2H+2},\label{A13}%
\end{equation}
\begin{equation}
\beta_{PL}(t)=\frac{1}{g}[\frac{g\tau (\alpha -2)+(1+g\tau )^{2-\alpha }-1}{\alpha -2}],
\end{equation}
where, $ H $ is known as the Hurst exponent and ranges as $ 0<H<1 $ \cite{57}. In the current case, the phase of the system is described by $\phi_n(t)=n \lambda \Omega_n(t)$. The dephasing effects of the noise are determined by averaging the time evolved density matrices over the corresponding noise phases as $\langle \exp[n \lambda \Omega_n(t)]\rangle = \langle \exp[-\frac{1}{2}n^2\beta_\mathcal{X_Y}(t)]\rangle$ where, $\mathcal{X_Y} \in \{FG, PL $\}. We get the final density matrix of the system within independent classical fields driven by the dephasing effects of the pure Gaussian noise case as:
\begin{equation}
\rho_p(\tau)= \left\langle \left\langle \left\langle U_{123}(t)\rho_o  U_{123}(t)^{\dagger} \right\rangle_{\theta_a} \right\rangle_{\theta_b} \right\rangle_{\theta_c}.\label{pure final density matrix}
\end{equation}
where, $\theta$ corresponds to the phase of either pure PL or FG noise. The final density matrix for the system of three qubits when coupled to independent classical fields in the PLM and FGM configurations are given by \cite{54}:
\begin{equation}
\rho_{PLM}(\tau)= \left\langle \left\langle \left\langle U_{123}(t)\rho_o  U_{123}(t)^{\dagger} \right\rangle_{\phi_a} \right\rangle_{\phi_b} \right\rangle_{\varphi_c},\label{2PL final density matrix}
\end{equation}
\begin{equation}
\rho_{FGM}(\tau)= \left\langle \left\langle \left\langle U_{123}(t)\rho_o  U_{123}(t)^{\dagger} \right\rangle_{\varphi_a} \right\rangle_{\varphi_b} \right\rangle_{\phi_c},\label{2FN final density matrix}
\end{equation}
where $\phi$ and $\varphi$ are the corresponding noises phases of the PL and FG noise. 
\section{Results and discussions}\label{Results and Discussions}
In this section, the analytical and numerical results obtained for the entanglement witness, purity, and decoherence measures for the dynamics of the three non-interacting qubits when coupled independently to external random fields generating pure and mixed classical noises are presented.
\subsection{Pure power-law and fractional Gaussian noise}\label{pure FGN and PLN}
\begin{figure}[ht]
\includegraphics[scale=0.09]{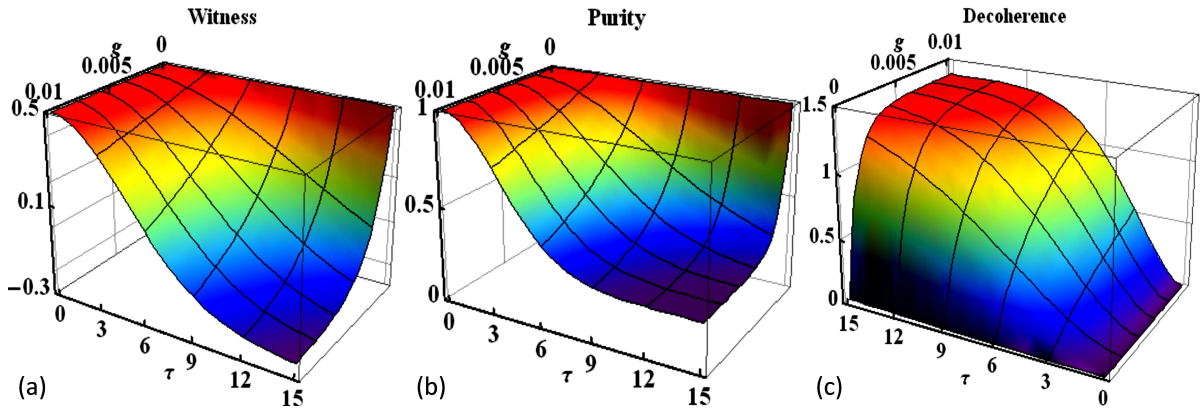}
\caption{Dynamics of the entanglement witness (a), purity (b) and decoherence (c) as a function of $\tau$ for the three non-interacting qubits, initially prepared in the states $\rho_{GHZ}(0)$ when subjected to independent classical fields with pure PL noise when $g=10^{-2}$ and $\alpha=2.1$.}\label{pure pln}
\end{figure}

\begin{figure}[ht]
\includegraphics[scale=0.09]{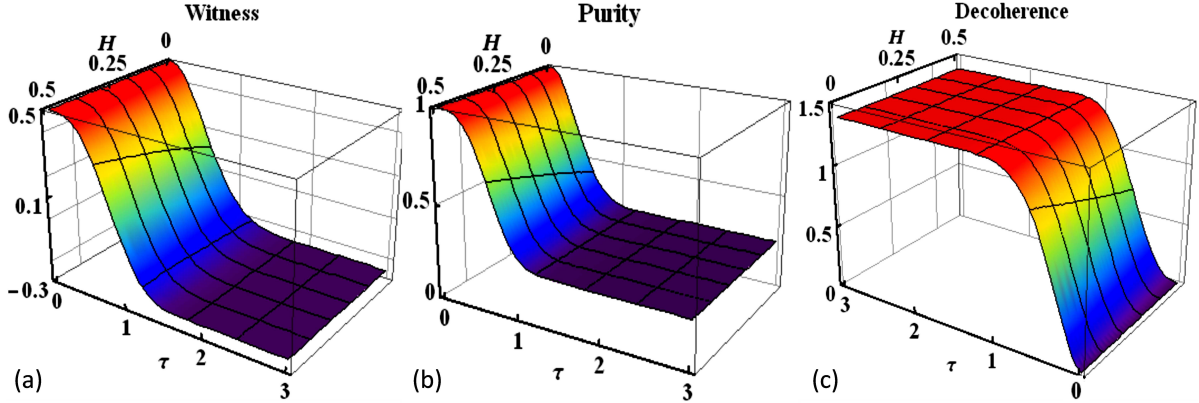}
\caption{Dynamics of the entanglement witness (a), purity (b) and decoherence (c) as a function of $\tau$ for the three non-interacting qubits, initially prepared in the states $\rho_{GHZ}(0)$ are subjected to independent classical fields with pure FG noise when $H=0.5$.}\label{pure FGN 3D}
\end{figure}

The analytical results obtained for the dynamics of entanglement, witness, purity and decoherence when coupled to classical independent environments with pure PL and FG noise are presented here. By using Eq.\eqref{pure final density matrix}, the final density matrix ensemble for the current pure noise case obtained is X-shaped, as shown in sect.\ref{Appendix}. This resembles the time evolved density ensemble state obtained for non-local-system environment coupling driven by random telegraph noise given in \cite{555}. All the diagonal elements are nonvanishing in the final density matrix to represent the three qubits to be in the chosen basis. Using Eqs.\eqref{ewo}, \eqref{purity} and \eqref{decoherence} for the final density matrix of the three-qubit state, the analytical expression are followed as (also shown in Figs.\ref{pure pln} and \ref{pure FGN 3D}):
\begin{align}
E_{p}(t)=&-\frac{1}{4}(1+3 \eta_1),\\
P_{p}(t)=&-\frac{1}{4}(1+3 \eta_2),\\
D_{p}(t)=&-\frac{3}{4} \mathcal{M} \log\left[\frac{1}{4} \mathcal{M} \right]-\frac{1}{4} \mathcal{N} \log\left[\frac{1}{4}  \mathcal{N} \right],
\end{align}
where $\eta_i=\exp[-n\beta_{\mathcal{XY}}]$ with $n \in \{4,8\}$.\\
Fig.\ref{pure pln} analyze the time evolution of the entanglement and coherence initially encoded in the tripartite entangled state when subjected to independent environments characterized by pure PL noise. Not all the off-diagonal elements are vanishing, thus represents the existence of the coherence between the qubits. Under the current dephasing effects, $E(t)$ and $P(t)$ were found decreasing functions of entanglement and coherence while the $D(t)$ remained an increasing function of coherence decay. The dynamical outlook appraises the dominant character of the PL noise to degrade the initial encoded entanglement and coherence in the system. We noticed that for the utmost low values of $g$, quantum correlations, as well as coherence, can be successfully preserved in the three-qubit systems, however, for finite lengthy intervals.
The random behaviour of the classical environments is completely suppressed upon the superposition of the noise phase over the joint phase of the system and environments. Because of this, the decay outlook turns out to be completely monotonic rather than showing any entanglement revivals. The backflow mechanism of information between the system and environment is thus vanished, resulting in permanent information loss. In contrast, temporary information losses have been observed under static, dynamic and coloured noise for different quantum systems \cite{37,38,53}. The coherence decay rate for the three-qubit system leads to the entanglement decay rate. This suggests that the decoherence between the system and the environment caused the system to be disentangled.\\
Fig.\ref{pure FGN 3D} evaluates dynamics of the entanglement and coherence initially encoded in a three-qubit state and subjected to pure FG noise in classical independent environments. The decay outlook in the current case differs from observed under pure PL noise. The pure FG noise seems more detrimental relatively, causing the three-qubit state to lose coherence and entanglement in a very short time as compared to the PL noise. The unusual character of the $H$ has been noticed to robust the entanglement and coherence which contradicts most of the noise parameters properties \cite{37,38, 48,53}. However, because of the dominant dephasing power of the current noise phase, even this robustness cannot prevent the state from becoming decoherent and disentangled. All other dynamical aspects match with those given in Fig.\ref{pure pln}.
\subsection{Power-law noise maximized configuration}\label{PLFN}
\begin{figure}[ht]
\includegraphics[scale=0.09]{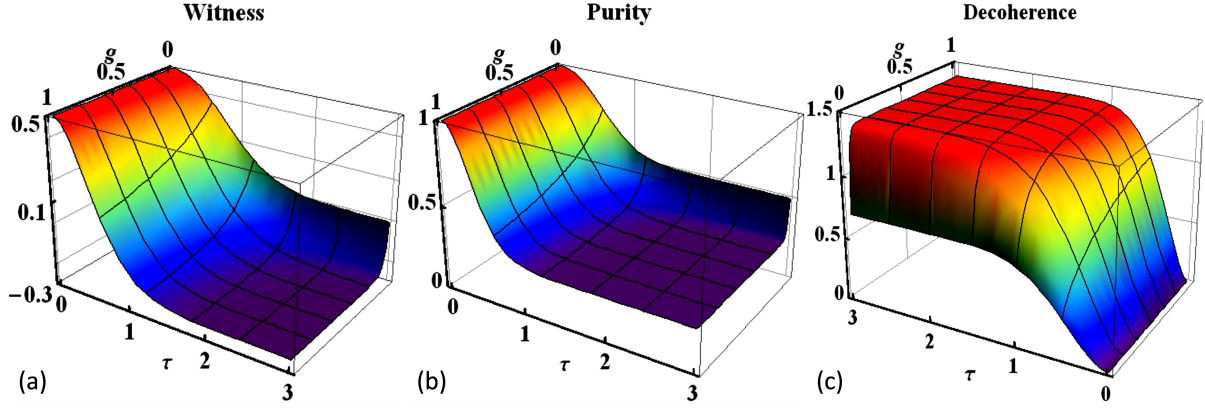}
\caption{Dynamics of the entanglement witness (a), purity (b) and decoherence (c) as a function of $\tau$ for the three non-interacting qubits, initially prepared in the states $\rho_{GHZ}(0)$ when subjected to independent classical fields for the case of power-law noise maximized configuration, when $H=0.1$, $\alpha=3$ and $g=1$.}\label{PLFN3D}
\end{figure}

The dynamics of entanglement witness, purity, and decoherence are evaluated for the PLM where two qubits are individually coupled with PL noise and one with FG noise. The final density matrix ensemble for the current mixed noise case has an X-shape, as shown in sect.\ref{Appendix}, and looks similar to the pure noise case. However, compared to the previous noise case, the density matrix elements and the phases differ. The structure of the current density matrix is like that obtained in \cite{555} for non-local system-environment coupling under random telegraph noise. In the final density matrix, all the diagonal elements are non-vanishing, indicating that the three qubits are in the chosen entangled basis. Besides, some non-vanishing elements of the final density matrix represent that the system is still coherent under the mixed noise situation. All the operations are done over the final density matrix given in Eq.\eqref{2PL final density matrix}. By using Eqs.\eqref{ewo}, \eqref{purity} and \eqref{decoherence}, the analytical results for the witness, purity and decoherence are followed as (also shown in Fig.\ref{PLFN3D}):
\begin{align}
E_{PLM}(t)=&\frac{1}{4} \left(-1+e^{-4 \beta_\mathcal{P_L}}+2 e^{-2 (\beta_{mix})}\right),
\end{align}
\begin{align}
P_{PLM}(t)=&\frac{1}{4} \left(1+e^{-8 \beta_\mathcal{P_L}}+2 e^{-4 (\beta_{mix})}\right),\\
D_{PLM}(t)=&-\frac{1}{2} P \log\left[\frac{1}{4} P\right]-\frac{1}{2} Q \log\left[\frac{1}{4} Q\right],
\end{align}
where, $\beta_{mix}$ represents the local-mixed noise phases of the PL and FG noise and is defined as $\beta_{mix}=\beta_{(\mathcal{F_G}+\mathcal{P_L})}$.\\
Fig.\ref{PLFN3D} shows that under the mixed noisy effects, witness and purity are both decreasing functions of entanglement, while decoherence is a function of increased coherence decay under PLM configuration. This shows the detrimental dephasing effects of the mixed noise phases on the tripartite entanglement and coherence dynamics. The decoherence and purity decay rates hit the saturation levels faster than the witness. It infers the coherence loss to be quicker than the entanglement. It means incoherency between the system and environment gives rise to the disentanglement between the three qubits. The rise in decoherence allows the qubits to be entangled partially with the surrounding environments that result in quantum information loss and separability. All the three measures display monotonic decay rather than showing any rebirths showing similarity with the decay caused by the Ornstein Uhlenbeck noise \cite{56,55}. This exponential decay apprises that the information lost by the system is not recovered back from the coupled classical channels and is permanently lost. The qualitative behavioural decay shown by the measures is likely the same and infer that, after a finite interaction time, the state becomes completely disentangled and decoherent. The slopes decay faster as the value of $g$ increases, indicating the dephasing nature of the parameter. We also find the current dephasing effects under the current mixed noisy situation completely different from that of the non-Gaussian noises, such as of random and static nature investigated in \cite{37,38,48,gg,ff}.
\subsection{Fractional Gaussian noise maximized configuration}
\begin{figure}[ht]
\includegraphics[scale=0.09]{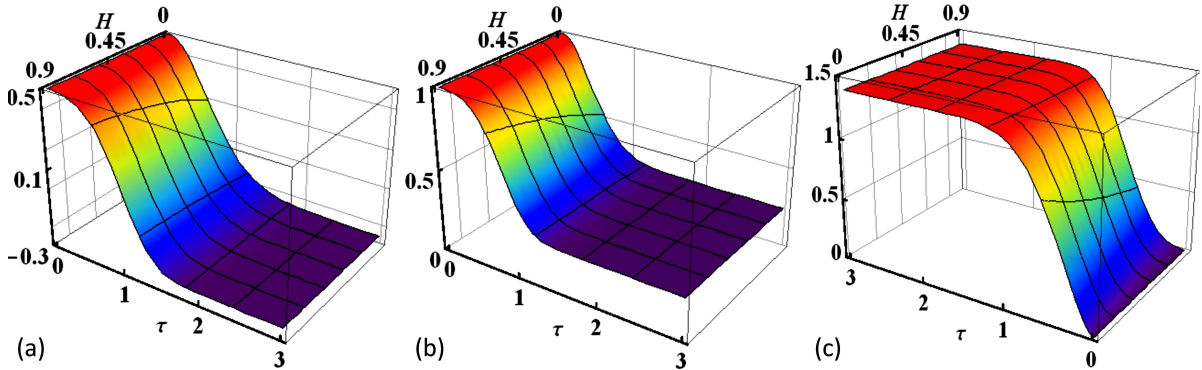}
\caption{Dynamics of the entanglement witness (a), purity (b) and decoherence (c) as a function of $\tau$ for the three non-interacting qubits, initially prepared in the states $\rho_{GHZ}(0)$ when subjected to independent classical fields for the case of fractional Gaussian noise maximized configuration, when $H=0.9$, $\alpha=3$ and $g=0.1$.}\label{FNPL3D}
\end{figure}

Here, we analyze the dynamics of entanglement witness, purity, and decoherence for three non-interacting maximally entangled qubits. Here, the independent system-environment coupling is assumed to be characterized by FGM, where two qubits are driven by FG and the third one by PL noise. The final density matrix state for the current mixed noisy situation has a similar shape and characteristics as the PLM, but with different noise phases. By using the Eqs.\eqref{ewo}, \eqref{purity} and \eqref{decoherence} for the final density matrix of the three-qubit state given in Eq.\eqref{2FN final density matrix}, the analytical results for the witness, purity and decoherence are followed as (also shown in Fig.\ref{FNPL3D}):
\begin{align}
E_{FGM}(t)=&\frac{1}{4} \left(-1+e^{-4 \beta_\mathcal{F_G}}+2 e^{-2 (\beta_{mix})}\right),\\
P_{FGM}(t)=&\frac{1}{4} \left(1+e^{-8 \beta_\mathcal{F_G}}+2 e^{-4 (\beta_{mix})}\right),\\
D_{FGM}(t)=&-\frac{1}{2} X \log\left[\frac{1}{4} X\right]-\frac{1}{2} Y \log\left[\frac{1}{4} Y\right].
\end{align}
Fig.\ref{FNPL3D} evaluates the dynamics of tripartite non-local correlation and coherence within independent classical channels driven by FGM configuration. In the current case, the quantitative and qualitative dynamical behaviour of the entanglement, coherence seems to be increasingly different from that observed under PLM configuration in Fig.\ref{PLFN3D}. For the increasing values of the Hurst index ($H$), the entanglement and coherence preservation becomes more robust initially, which completely contradicts the previous results for the noise parameter $g$ of the PL noise. The dynamics of entanglement and coherence are not only affected by the coupled classical environments but is also significantly altered by the number of qubits driven by a specific noise and the values of the corresponding noise parameters. The behaviour of the noise phases over the qubit sub-spaces caused the initially encoded non-local correlation and coherence to degrade. Under the presence of mixed noisy phases, both witness and purity are found to be decreasing functions of the entanglement and coherence. In contrast, the decoherence measure is found to be the increasing function of coherence decay. However, the rates of entanglement decay and coherence decay differ slightly in time, with entanglement reaching the final saturation levels later than the coherence. This implies that the disentanglement of quantum systems highly depends on the amount of decoherence existence between the system and its environment. This suggest the coherence to be a necessary prerequisite for the entanglement preservation. As can be seen, the disentanglement rate directly increases with the increasing speed of the decoherence. The three measures are monotonous functions in time, with no evidence of the entanglement of sudden death and birth phenomena. This resembles the bipartite and tripartite entanglement decay observed in \cite{56,55,ff} due to the Ornstein Uhlenbeck noise. However, the quantitative aspects are completely different. This kind of monotonic decay results in the off-limitation of the back-flow of information from the environment to the tripartite states after it has been lost. After a finite interaction time, the measures show that the three qubits become separable, mixed, and decoherent. Thus, all three measures show the same qualitative dynamical behaviour, showing a firm agreement among them.
\subsection{Detailed time evolution analysis}
In this section, we provide a detailed analysis including comparative dynamics, memory properties, and detrimental effects of the pure PL and FG noise along with the mixed Gaussian noise cases, PLM and FGM configurations. The focus of the section is to estimate the dephasing effects explicitly against different values of the corresponding noise parameters in the given pure and mixed noisy configurations.
\subsubsection{The case of comparing the dephasing effects of the pure power-law and fractional Gaussian noise}
When coupled to independent classical fluctuating fields generating pure PL noise (red-lined slopes) and FG (blue-lined slopes), the time evolution of entanglement witness, purity, and decoherence for the tripartite $\mathcal{X_{GHZ}}$ state is described. The current section relates to the ability of the two pure Gaussian noises to degrade entanglement and coherence in terms of the noise parameters $g$ and $H$.\\
\begin{figure}[ht]
\includegraphics[scale=0.09]{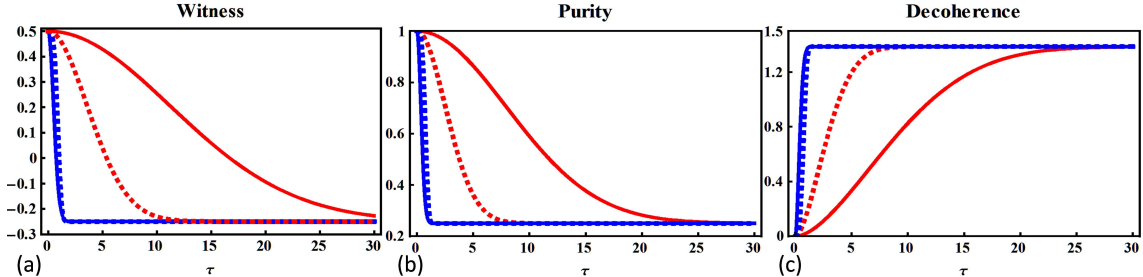}
\caption{Dynamics of the entanglement witness (a), purity (b) and decoherence (c) as a function of $\tau$ for the three non-interacting qubits, initially prepared in the states $\rho_{GHZ}(0)$ when subjected to independent classical fields, with pure PL (red-lines slopes) and FG noise (blues-lined slopes) when $H=10^{-2}$, $g=10^{-3}$, $\alpha=2.1$ (non-dashed slopes) $H=0.9$ and $g=10^{-1}$, $\alpha=3$ (dashed-slopes).}\label{Comparison of the pure noises}
\end{figure}
Fig.\ref{Comparison of the pure noises} analyses the dynamics of time-dependent entanglement witness, purity and decoherence for tripartite states within independent classical fields driven by pure FG and pure PL noise. The difference between the resulted dynamics of the entanglement and coherence is very much distinguishable under the two different Gaussian noises. We noticed that the entanglement and coherence remain much short-lived under FG noise. Besides, we found the phase of the PL noise readily exploitable to get long-ranged quantum correlations and coherence preservation. The spectrum of the FG noise seems much narrower than that of the PL noise. Following this, one can note that the variation in the preservation time of the FG noise is negligible compared to that shown by the opponent noise. The statement can be verified by comparing the graphical results given in Fig.\ref{pure FGN 3D} and Fig.\ref{Comparison of the pure noises}. Other qualitative dynamical properties of the three qubits are in good agreement with those defined in Fig.\ref{pure FGN 3D} and \ref{PLFN3D}. However, with PL noise, the preservation interval differs significantly. The current long interval preservation of the non-local correlation and coherence is completely attributable to the utmost low values of the noise parameter $g$. Thus, making it easier for the quantum practitioners to design long-range entanglement, coherence and memory properties for the successful deployment of the quantum mechanical protocols. In comparison, the preservation effects under FG noise are much lesser than those obtained under Ornstein Uhlenbeck noise \cite{55,56}, but much longer preserved entanglement and coherence has been obtained under the current PL noise.
\subsubsection{The case of comparing the dephasing effects of the power-law and fractional Gaussian noise maximized configurations}
Time evolution of the entanglement, purity, and coherence of the tripartite $\mathcal{X}_{GHZ}$ state, when coupled to independent classical fluctuating fields generating noises in the PLM and FGM configurations is addressed here. The current section focuses on the potential of the two mixed noisy configurations to degrade entanglement and coherence in comparison. In the first case, we investigate the detrimental effects of the mixed-noise configuration when two qubits are coupled to the independent classical fields with PL noise and the third to a field with FG noise (as shown by the red-lined slopes). In the later noisy configuration, two qubits are under the influence of FG noise and one under the PL noise (as shown by the blue-lined slopes).\\
\begin{figure}[ht]
\includegraphics[scale=0.09]{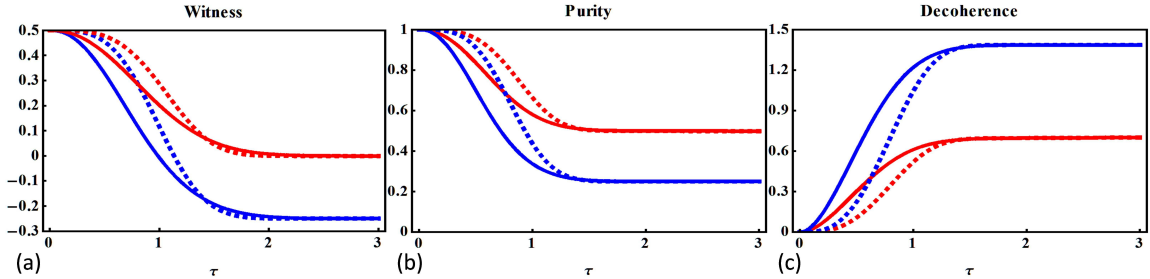}
\caption{Dynamics of the entanglement witness (a), purity (b) and decoherence (c) as a function of $\tau$ for the three non-interacting qubits, initially prepared in the states $\rho_{GHZ}(0)$ when subjected to independent classical fields under power-law noise maximized configuration (red-lined slopes) and fractional Gaussian noise maximized configuration (blue-lined slopes) when $H=10^{-1}$, $\alpha=2.1$ (non-dashed slopes) and $H=0.8$, $\alpha=10$ (dashed slopes) with $g=10^{-4}$.}\label{Comparison of the mixed noises}
\end{figure}
In Fig.\ref{Comparison of the mixed noises}, the time evolution of the entanglement witness, purity, and decoherence under the dephasing effects of two different mixed-noisy configurations is reported. The current results show that both noisy schemes dephase the initially maximally entangled tripartite state, causing it to separable and decohere after a finite interaction time. The dephasing effects caused by these two mixed noisy schemes differ significantly. As shown, when two of the qubits were coupled to FG noise, the initially encoded entanglement and coherence suffered a greater loss. This suggests that the FG noise phase is more disruptive than the PL noise phase, which produced smaller disentanglement and decoherence effects. As a result, the nature of decay can be deduced to largely depend on the type of noise involved. In Figs.\ref{Comparison of the pure noises} and \ref{Comparison of the mixed noises}, the net difference between the preservation effects is significantly large. Entanglement and coherence are preserved for a longer time under pure PL noise, but not for long enough under both mixed noisy configurations. The main reason for modelling long entanglement and coherence preservation is the easily exploitable noise phase of the PL noise. However, it is difficult to avoid mixed noisy dephasing effects, especially when FG noise is involved, resulting in shorter memory effects. The dominant dephasing effects resulting in faster entanglement and coherence decay are easily deducible under FG noise in both pure and mixed noisy configurations, as shown in Figs.\ref{pure FGN 3D} and \ref{Comparison of the mixed noises}. In comparison, the current results under both mixed noisy configurations differ completely from those defined in \cite{JSR} for joint dephasing effects of static and random telegraph noise. In addition, the qualitative dynamical behaviour of the entanglement, purity, and decoherence coincides with one another and with previous findings. The decay observed is fully monotonic, according to the measures, and no entanglement sudden death and birth revivals have been observed. We showed that, in addition to the noisy character of the parameter $g$, $\alpha$ has the dephasing nature too. Entanglement and coherence decay increase in direct proportion to the value of $\alpha$. As $H$ increases, the entanglement and coherence become more robust at first, but the total preservation duration remains unaffected. The purity and coherence decay rates have been observed to be faster than the disentanglement rates of the three qubits, which are, however, directly related. Excluding the decay rates, all the three measures showed consistent results, indicating good agreement among them. The current detrimental effects resemble the decay caused by the Ornstein Uhlenbeck noise \cite{56,55,ff} but strongly disagree with the dephasing effects because of non-Gaussian noises given in \cite{37,38,48,53}.
\subsubsection{Power-law noise maximized configuration: entanglement and coherence dynamics under the wide range of the noise parameter $g$}
The time evolution of entanglement witness, purity, and coherence for three non-interacting qubits when coupled independently to external fields generating mixed Gaussian noises is presented. Here, we assume the two of the independent fields have PL noise and one has FG noise, with one qubit coupled to each. In this context, we focus on the ability of the current mixed noise case to destroy the entanglement and coherence because of different noise parameter values.\\
\begin{figure}[ht]
\includegraphics[scale=0.09]{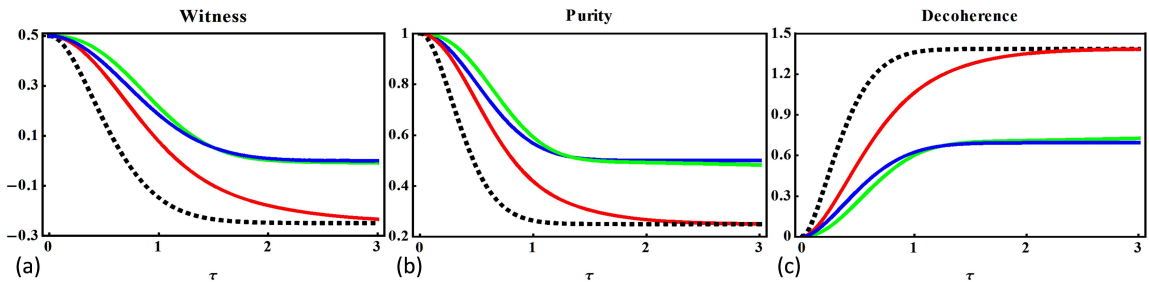}
\caption{Dynamics of the entanglement witness (a), purity (b) and decoherence (c) as a function of $\tau$ for the three non-interacting qubits, initially prepared in the states $\rho_{GHZ}(0)$ when subjected to independent classical fields under power-law noise maximized configuration when $g=10^{-2}$ (blue), $10^{-1}$ (green), $1$ (red) and $10$ black with $\alpha=3$ and $H=10^{-1}$.}\label{2PL-FN different noise parameters}
\end{figure}
Fig.\ref{2PL-FN different noise parameters} shows the time evolution of entanglement witness, purity, and decoherence against the mixed Gaussian noises originated from the independent coupling of the three qubits to classical fields. We found that the current mixed noise configuration has different adverse effects than the case of the maximized effects of FG noise. The spectrum of the PL noise is not as discrete as that of the FG noise, encompassing a variety of decay. Here, the deterioration of entanglement and coherence increases directly with the increasing values of $g$. The shifting of the blue-lined slopes follows this towards the red end for the increasing values of $g$. Furthermore, $g$ controls not only the decay rate but also the decay levels. As seen, the decay accumulates smaller values for smaller values of the parameter, while the decay levels rise for higher values of $g$. In agreement with the previous results, the nature of the decay encountered is completely monotonous functions in time and no entanglement sudden death and birth revivals were observed which resembles the dephasing effects in tripartite entanglement and coherence caused by Ornstein Uhlenbeck noise \cite{54,55}. This explains the quantum information's irreversible decay because of the present mixed-noise configuration's detrimental effects. However, under the non-Gaussian noises, the overall dynamical behaviour of the current tripartite state becomes increasingly different, as shown in \cite{37, 38, rr, ff, ii,jj}. The FG noisy effects over the third qubit are reduced in the PLM configuration, showing tiny evidence of presence. As a result, the nature of the noise and its application to the number of qubits play a pivotal role in determining the kind, decay amount and preservation interval. In the current mixed noise case, even for smaller values of the noise parameters, the disentanglement and decoherence effects are unavoidable. However, the decay at the smaller values of $g$ is relatively much smaller than that at the higher values. All three measures produced the same qualitative results, ensuring higher consistency and validity in the results. The disentanglement rate shown by the entanglement witness occurs later in the qualitative scales than the coherence decay rate shown by the purity and decoherence measure. This shows that the generation of decoherence causes that disentanglement between the system and the environment and that it increases in direct relation to the rate of coherence decay.
\subsubsection{Fractional Gaussian noise maximized configuration: entanglement and coherence dynamics under the discrete nature of the parameter $H$.}
In this section, we investigate the time evolution of entanglement witness, purity, and coherence for the $\mathcal{X}_{GHZ}$ state under the influence of FGM configuration. In the current case, the first two qubits are driven by FG noise and one by PL noise. Here, the preservation of the non-local correlation and coherence against the different values of the noise parameter $H$ is primarily investigated.\\
\begin{figure}[ht]
\includegraphics[scale=0.09]{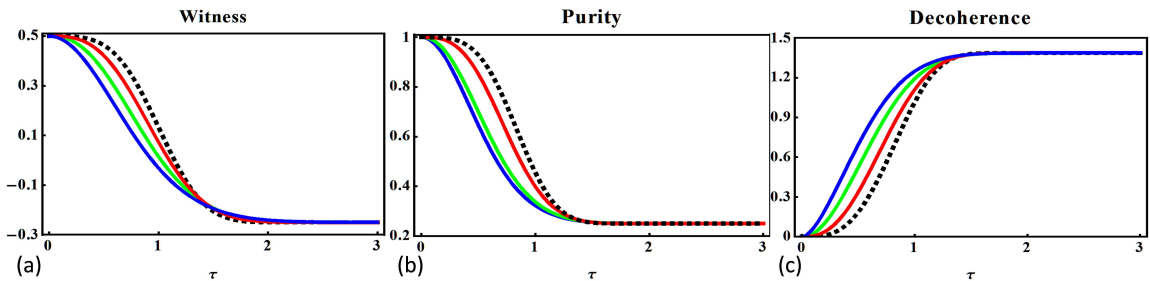}
\caption{Dynamics of the entanglement witness (a), purity (b) and decoherence (c) as a function of $\tau$ for the three non-interacting qubits, initially prepared in the states $\rho_{GHZ}(0)$ when subjected to independent classical fields under fractional Gaussian maximized configuration when $H=10^{-2}$ (blue), $0.2$ (green), $0.5$ (red) and $0.9$ black with $\alpha=3$ and $g=10^{-3}$.}\label{2FN-PL different noise parameters}
\end{figure}
Fig.\ref{2FN-PL different noise parameters} evaluates the dynamical behaviour of tripartite entanglement and coherence when subjected to independent environments with joint effects of FG and PL noise. Unlike the PL noise, the width between the slopes for different values of $H$ is much smaller in comparison, which is because of the discrete nature of the FG noise with a narrow spectrum. For increasing values of $H$, the non-local correlation and coherence remained more robust initially, contrary to the characteristic behaviour of $ g$. As with increasing choices of $H$, the slopes shift from blue towards the black-end, indicating later decay. The dephasing effects caused by the FG noise, on the other hand, are inevitable, and the state becomes completely disentangled and decoherent after a very short interaction period. The increased effects of the FG noise over the two qubits suppress the PL noisy effects over the third qubit to the point that they do not even show up. It should be noted, however, that the PL noise parameters $g$ and $\alpha$ are kept minimal. The nature of the decay was unaffected by the increasing values of the noise parameters, remaining monotonic and free of entanglement sudden death, and birth revivals. This agrees with the previous results obtained for Ornstein Uhlenbeck noise \cite{54,55,56}, however, long-lived correlations and coherence with strong entanglement rebirths have been observed under static, random telegraph and coloured noises comparatively \cite{37, 38, rr, ff, ii,jj}. Besides, the observed disentanglement rate is slower than the purity and coherence decay rates, which is consistent with previous findings. The measures have shown similar qualitative dynamical behaviour, implying close connections between them.
\section{Conclusion}\label{Conclusion}
The dynamics of tripartite entanglement and coherence for a system of three non-interacting qubits prepared as maximally entangled GHZ-like state coupled with independent classical environments are examined in a brief description. The external fields are considered being characterized by four different noisy models: pure power-law, pure fractional Gaussian, power-law noise maximized, and fractional Gaussian noise maximized configurations. We computed the final density matrices for the three qubits by taking ensemble averages over the stochastic process in both the pure and mixed independent environmental noisy configurations. Finally, using estimators such as entanglement witness, purity, and decoherence, we probed the dynamics and preservation of entanglement and coherence in three qubits.\\
Our results show that the dephasing effects on the dynamics of tripartite non-local correlation and coherence are fundamentally different in pure and mixed noisy configurations. The PL noise was more damaging in the mixed noisy condition than in the pure noisy case. The related phase in the pure PL noise is easily exploitable to induce entanglement and coherence preservation for a long extended time, as shown in Fig.\ref{Comparison of the pure noises}. In contrast, FG noise is equally destructive in both pure and mixed noisy contexts, causing entanglement and coherence to vanish in a very short time. It's also worth noting that in the mixed noisy configuration, the dephasing effects of noise applied to a single qubit are completely suppressed, with only an infinitesimal evidence of presence. As a result, the type of noise applied to the corresponding number of qubits has been observed to affect both the decay rate and amounts. Besides this, the decay amounts are smaller when two qubits are coupled with PL noise and greater when two qubits are coupled with FG noise, as shown in Fig.\ref{Comparison of the mixed noises}. In addition, both pure and mixed noisy effects within independent classical fields are unavoidable, and with either a long or short preservation period, the state eventually becomes separable. Moreover, we found the tripartite and bipartite quantum correlations dynamics under the mixed non-Gaussian noisy configuration given in \cite{ff, JSR} completely different than the current Gaussian mixed noise cases. The basic difference between the two is the presence of the entanglement revivals, due to which, unlike the current case, the state remained inseparable for longer intervals.\\
The decay was monotonic, and there was no sign of entanglement sudden death and birth phenomenon, which contradicts the findings in \cite{37,38,48,53,gg, ff,ii,jj} for various non-Gaussian noises. This signifies that entanglement and coherence, as well as quantum information degradation, are irreversible once lost, as also observed under Ornstein Uhlenbeck noise in \cite{54,55,56}.\\
Entanglement, coherence, and memory features were initially more robust for the upper bound of the $H$ in the case of noise parameters, in contrast to the characteristics of most of the noisy parameters described in \cite{gg,53,55,56,55,ff,48}. The preservation intervals, on the other hand, were almost unchanged across the whole discrete range of the $H$. Compared to the broad range of $g$ of the PL noise, the variation in decay induced by the FG noise is nearly trivial over the entire range of the parameter $H$. This characteristic of the FG noise is fully attributable to the discrete spectrum of the noise. In the case of power-law noise, the parameters $g$ and $\alpha$ play opposing roles in the dynamics of non-local correlation and coherence, with both decreasing as this parameter is increased. Aside from that, due to the broad spectrum of the power-law noise, a variety of decay has been seen against various values of $g$. Most significantly, much longer and greater entanglement and coherence retention may be achieved for low values of $g$, far longer than those obtained under Gaussian and non-Gaussian noises investigated in \cite{ii,jj,ff,37,38,53,55,54}\\
Equal qualitative dynamical behaviour was seen in the case of the measures employed, suggesting a strict agreement between them and ensuring consistency and validity in the results. In quantitative analysis, the purity and decoherence measures, on the other hand, exhibited a quicker deterioration rate than the entanglement witness. This means that the rate of disentanglement is proportionally caused by the amount of decoherence between the system and its environments.\\
Finally, we find that the current Gaussian noisy effects within classical independent fields are inescapable in any situation. For tripartite entanglement, coherence and memory characteristics, we find fractional Gaussian noise to be more detrimental than power-law noise. The pure power-law noise assisted local channels are detected with the least dephasing effects, allowing the survival of non-local correlation and coherence for longer intervals, notably for the extremely low values of the parameters $g$ and $\alpha$. We found that no such exploitation is conceivable in the case of fractional Gaussian noise that can mimic long enough preservation effects.
\section{Appendix}\label{Appendix}
In this section, we give the details of the final density matrices obtained for the time evolution of the three qubits initially prepared in the state $\rho_{GHZ}(0)$ under the effects of pure and mixed noisy configurations. Using the Eq.\eqref{pure final density matrix}, we get the final density matrix under pure Gaussian noise case as:
\begin{align}
\rho_p(t)=\frac{1}{8}\left[
\begin{array}{cccccccc}
 1+3 \mathcal{X}_1 & 0 & 0 & 0 & 0 & 0 & 0 & 1+3 \mathcal{X}_1 \\
 0 & 1-\mathcal{X}_1 & 0 & 0 & 0 & 0 & 1-\mathcal{X}_1 & 0 \\
 0 & 0 & 1-\mathcal{X}_1 & 0 & 0 & 1-\mathcal{X}_1 & 0 & 0 \\
 0 & 0 & 0 & 1-\mathcal{X}_1 & 1-\mathcal{X}_1 & 0 & 0 & 0 \\
 0 & 0 & 0 & 1-\mathcal{X}_1 & 1-\mathcal{X}_1 & 0 & 0 & 0 \\
 0 & 0 & 1-\mathcal{X}_1 & 0 & 0 & 1-\mathcal{X}_1 & 0 & 0 \\
 0 & 1-\mathcal{X}_1 & 0 & 0 & 0 & 0 & 1-\mathcal{X}_1 & 0 \\
 1+3 \mathcal{X}_1 & 0 & 0 & 0 & 0 & 0 & 0 & 1+3 \mathcal{X}_1
\end{array}
\right]
\end{align}
Next, using the Eq,\eqref{2FN final density matrix}, we obtained the final density matrix ensemble state as:
\begin{align}
\rho_{XYZ}(t)=\frac{1}{8}\left[
\begin{array}{cccccccc}
\mathcal{H}_1 & 0 & 0 & 0 & 0 & 0 & 0 &\mathcal{H}_1 \\
 0 & \mathcal{H}_2 & 0 & 0 & 0 & 0 & \mathcal{H}_2 & 0 \\
 0 & 0 & \mathcal{H}_3 & 0 & 0 & \mathcal{H}_3 & 0 & 0 \\
 0 & 0 & 0 & \mathcal{H}_3 & \mathcal{H}_3 & 0 & 0 & 0 \\
 0 & 0 & 0 & \mathcal{H}_3 & \mathcal{H}_3 & 0 & 0 & 0 \\
 0 & 0 & \mathcal{H}_3 & 0 & 0 & \mathcal{H}_3 & 0 & 0 \\
 0 & \mathcal{H}_2 & 0 & 0 & 0 & 0 & \mathcal{H}_2 & 0 \\
\mathcal{H}_1 & 0 & 0 & 0 & 0 & 0 & 0 &\mathcal{H}_1
\end{array}
\right]
\end{align}
where $\rho_{XYZ}(t) \in \{ \rho_{PLM}(t), \rho_{FGM}(t)\}$, $\mathcal{X}_i=\exp[-n\beta_{\mathcal{A_B}}(t)]$ with $\beta_{\mathcal{A_B}}(t) \in \beta_{\mathcal{P_L}}(t), \beta_{\mathcal{F_G}}(t)$ and $\mathcal{Y}=\exp[-n\beta_{mix}(t)]$ with $\beta_{mix}(t)=\beta_{\mathcal{P_L}}(t)+\beta_{\mathcal{F_G}}(t)$, $\mathcal{H}_1=1+\mathcal{X}_2+2 \mathcal{Y}$, $\mathcal{H}_2=1+\mathcal{X}_2-2 \mathcal{Y}$, $\mathcal{H}_3=1-\mathcal{X}_2$.
\section{Data Availability}
The authors confirm that the data supporting the findings of this study are available within the article and its supplementary materials.

\end{document}